\documentclass[%
 reprint,
superscriptaddress,
 amsmath,amssymb,
 aps,
 twocolumn,
 longbibliography,
]{revtex4-2}
\usepackage{tikz}

\usepackage{graphicx}
\usepackage{dcolumn}
\usepackage{bm}

\usepackage{tabularx}
\usepackage{flafter}
\usepackage{wasysym}
\usepackage{xcolor}
\usepackage{ulem}

\definecolor{darkgreen}{rgb}{0.0, 0.5, 0.0}

\begin{document}


\title{
Segregation forces in dense granular flows: Closing the gap from single intruders to mixtures
}

\author{Yifei Duan}
\affiliation{
Department of Chemical and Biological Engineering, Northwestern University, Evanston, Illinois 60208, USA \looseness=-1
}%
\author{Lu Jing}
\affiliation{
Department of Chemical and Biological Engineering, Northwestern University, Evanston, Illinois 60208, USA \looseness=-1
}%
\author{Paul B. Umbanhowar}
\affiliation{
Department of Mechanical Engineering, Northwestern University, Evanston, Illinois 60208, USA
}%
\author{Julio M. Ottino}
\affiliation{
Department of Chemical and Biological Engineering, Northwestern University, Evanston, Illinois 60208, USA \looseness=-1
}%
\affiliation{
Department of Mechanical Engineering, Northwestern University, Evanston, Illinois 60208, USA
}%
\affiliation{
Northwestern Institute on Complex Systems (NICO), Northwestern University, Evanston, Illinois 60208, USA \looseness=-1
}%
\author{Richard M. Lueptow}
 \email{Corresponding author: r-lueptow@northwestern.edu}
 \affiliation{
Department of Chemical and Biological Engineering, Northwestern University, Evanston, Illinois 60208, USA \looseness=-1
}%
\affiliation{
Department of Mechanical Engineering, Northwestern University, Evanston, Illinois 60208, USA
}%
\affiliation{
Northwestern Institute on Complex Systems (NICO), Northwestern University, Evanston, Illinois 60208, USA \looseness=-1
}%

\date{\today}

\begin{abstract}
Using simulations and a virtual-spring-based approach, we measure the segregation force, $F_\mathrm{seg},$ over a range of size-bidisperse mixture concentrations, particle size ratios, and shear rates to develop a model for $F_\mathrm{seg}$ that extends its applicability from the well-studied non-interacting intruders regime to finite-concentration mixtures where cooperative phenomena occur. The model predicts the concentration below which the single intruder assumption applies and provides an accurate description of the pressure partitioning between species.
\end{abstract}

\maketitle



Flowing granular materials tend to segregate by particle size, density, or other physical properties, which is a phenomenon crucial to many industrial and geophysical processes \cite{ottino2000mixing,ottino2008mixing,frey2009river}. Despite decades of research on this topic, fundamental aspects of granular flow-driven segregation remain elusive, and state-of-the-art continuum segregation models largely rely instead on {\it ad hoc} or configuration-specific closure schemes~\citep{gray2018particle,umbanhowar2019modeling,thornton2021brief}. Recent efforts characterizing forces on single intruder particles in otherwise species-monodisperse granular flows have advanced our understanding of segregation at the particle level~\citep{tripathi2011numerical,guillard2016scaling,jing2017micromechanical,van2018segregation,staron2018rising,jing2020rising} and led to segregation force models applicable across flow configurations \citep{guillard2016scaling,jing2021unified}. However, it is unclear whether or how single intruder results can be applied to granular mixtures with finite species concentration \citep{tripathi2021theory,rousseau2021bridging}. 
More fundamentally, the physical mechanisms governing transitions in segregation behaviors between intruder and mixture regimes as the species concentration is varied, remain unresolved.

\begin{figure}
    \centerline{\includegraphics[width=3.5 in]{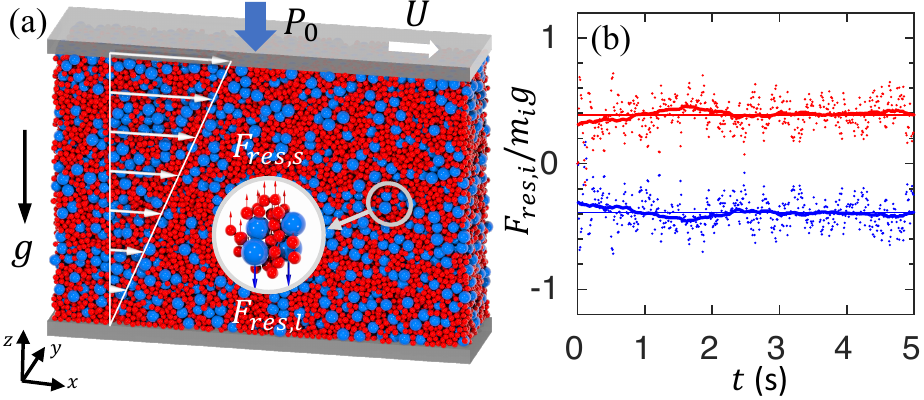}}
    \caption{(a) Large (4\,mm, {\color{blue}{blue}}) and small (2\,mm, {\color{red}{red}}) particles ($c_l=c_s=0.5$) in a controlled, constant-shear-rate flow.
(b) Scaled restoring force vs.\ time for large (\textcolor{blue}{blue}) and small (\textcolor{red}{red}) particles. Data points sampled at 0.01\,s intervals; bold curves are averages using a 1\,s long sliding window. Horizontal lines are averages from 2\,s to 5\,s.
    }
    \label{scheme}
\end{figure}

In this Letter, we show that particle size segregation in sheared granular flow exhibits a continuous transition from the single intruder limit to finite mixture concentrations. To do so, we extend the virtual-spring-based ``force meter'' approach for a single intruder particle~\citep{guillard2016scaling,van2018segregation,jing2020rising} to size-bidisperse mixtures of arbitrary species concentration and use it to characterize the dependence of the segregation force on concentration for various particle size ratios in controlled, constant-shear-rate flow simulations, see Fig.~\ref{scheme}(a). We find that the segregation force exhibits a plateau at small concentrations and decreases monotonically above a critical concentration, indicating a transition from non-interacting intruders to cooperative phenomena in mixtures, which is reminiscent of previously observed asymmetric concentration dependence in the segregation flux \cite{van2015underlying,jones2018asymmetric}. We also show that these results can provide physics-based closures for connecting segregation models with continuum theories for granular mixtures.

{\it Methods}. An in-house discrete element method (DEM) code running on CUDA-enabled GPUs~\cite{isner2020axisymmetric} is used to simulate a size-bidisperse particle mixture with volume concentration $c_i$, diameter $d_i$, and density $\rho_i=1$\,g/cm$^3$ ($i = l$, $s$ for large or small particles, respectively) sheared in a streamwise ($x$) and spanwise ($y$) periodic domain of length $L=35d_l$, width $W=10d_l$, and height $H=25d_l$ to $40d_l$ (varied as needed) in the presence of gravity ($g=9.81\,$m/s$^2$, in the negative $z$-direction), see Fig.~\ref{scheme}(a). The standard linear spring-dashpot model \cite{cundall1979discrete} is used to resolve particle-particle and particle-wall contacts of spherical particles using a friction coefficient of 0.5, a restitution coefficient of 0.2, and a binary collision time of 0.15\,ms. Changing bounding walls from smooth to bumpy (randomly attached particles) does not affect the results. Large ($d_l=4$\,mm) and small particles ($d_s$ varied to adjust the size ratio, $d_l/d_s$) have a $\pm10$\% uniform size distribution to minimize layering \cite{staron2014segregation} (increasing the size variation to $\pm20$\% does not alter the results).  

A constant shear rate $\dot\gamma=U/H$ is imposed on the flow by applying a streamwise stabilizing force, $F_{\mathrm{stabilize},k}=K_s (u_k-\dot\gamma z_k)$, on each particle $k$ at every simulation time step, where $u_k$ is the particle streamwise velocity, $z_k$ is the vertical particle position, and $K_s$ is a gain parameter~\cite{Lerner2012unified,clark2018critical,fry2018effect,saitoh2019nonlocal,duan2020segregation,jing2020rising}.  This stabilizing force reduces the granular temperature in the streamwise direction but does not alter the segregation~\cite{jing2021unified}.
An overburden pressure equal to the pressure at a depth of $H_w=20d_l$ (i.e., $P_{\mathrm{wall}}=\rho \phi g H_w$ where the bulk solid fraction $\phi$ varies from 0.55 to 0.6 depending on flow conditions) is applied using a massive flat frictional top wall that is free to move vertically (fluctuates by $\pm2\%$ or less after an initial rapid dilation of the particles at flow onset) and moves horizontally at a velocity determined by the constant shear rate velocity profile.

A spring-like vertical restoring force proportional to the center of mass distance between the two species is applied uniformly to all particles of each species $i$ at every simulation time step in order to characterize the particle forces while preventing segregation and the resulting changes in local concentration. This method is inspired by the virtual spring-based force technique used in single intruder DEM simulations to measure the segregation force \cite{guillard2016scaling,van2018segregation,jing2020rising}. The restoring force is $F_{\mathrm{res},i}=-K_r( \bar z_{i}-\bar z_j)/N_i$, where $\bar z_{i}={\sum_{k\in i}^{N_i} z_kV_k}/{\sum_{k= 1}^{N} V_k}$, $V_k$ is the volume of particle $k$, subscript $j$ indicates the other species, and $N_i$ and $N$ are the number of particles of species $i$ and the total number of particles, respectively. Since the applied restoring forces are internal forces,
\begin{equation}
F_{\mathrm{res},i} N_i+F_{\mathrm{res},j}N_j=0
\label{eq1}
\end{equation} 
and the bulk flow behavior (e.g., shear flow, bulk pressure) is  unaltered.

Figure~\ref{scheme}(b) plots the instantaneous restoring force scaled by particle weight, $F_{\mathrm{res},i}/m_ig$, at 0.01\,s intervals.
The scaled restoring forces for large (blue) and small (red) particles are equal and opposite for $c_l=c_s=0.5$ due to force balance, which can be written as $c_lF_{\mathrm{res},l}/m_lg+c_sF_{\mathrm{res},s}/m_sg=0$ based on Eq.~(\ref{eq1}), noting that particle mass $m_i=\rho V_i$ and species volume concentration $c_i=N_i V_i/V_{\mathrm{tot}}$, where $V_{\mathrm{tot}}$ is the total particle volume. The time average $F_{\mathrm{res},i}/m_ig$ over 1\,s time windows (bold curve) remains relatively constant 2\,s after flow onset, although force fluctuations occur due to the stochastic nature of granular flows. In addition, varying the shear rate $\dot \gamma$, the layer thickness $H$, or the gain parameters $K_s$ and $K_r$ has minimal influence on $F_{\mathrm{res},i}/m_ig$, indicating that the restoring force is independent of the details of the flow geometry and control parameters, and that its effect is uniform through the depth of the particle bed.

Since $F_{\mathrm{res},i}$, determined as the time-average of the reactive restoring force, balances the particle segregation force, $F_{\mathrm{seg},i}$ and the particle weight, $m_ig$,
\begin{equation}
F_{\mathrm{seg},i}=m_ig-F_{\mathrm{res},i}.
\label{equilibrium} 
\end{equation}
$F_{\mathrm{seg},i}$ is always upward, opposing gravity.  Since $F_{\mathrm{res},s}>0$ [Fig.~\ref{scheme}(b)], $F_{seg,s}<m_s g$ so small particles would sink without the restoring force; likewise, since $F_{\mathrm{res},l}<0$, $F_{seg,l}>m_l g$ so large particles would rise without the restoring force. Hereon, we scale the segregation force with the particle weight, $\hat F_i=F_{\mathrm{seg},i}/ m_i g$.

\begin{figure}
     \centerline{\includegraphics[width=3.5 in]{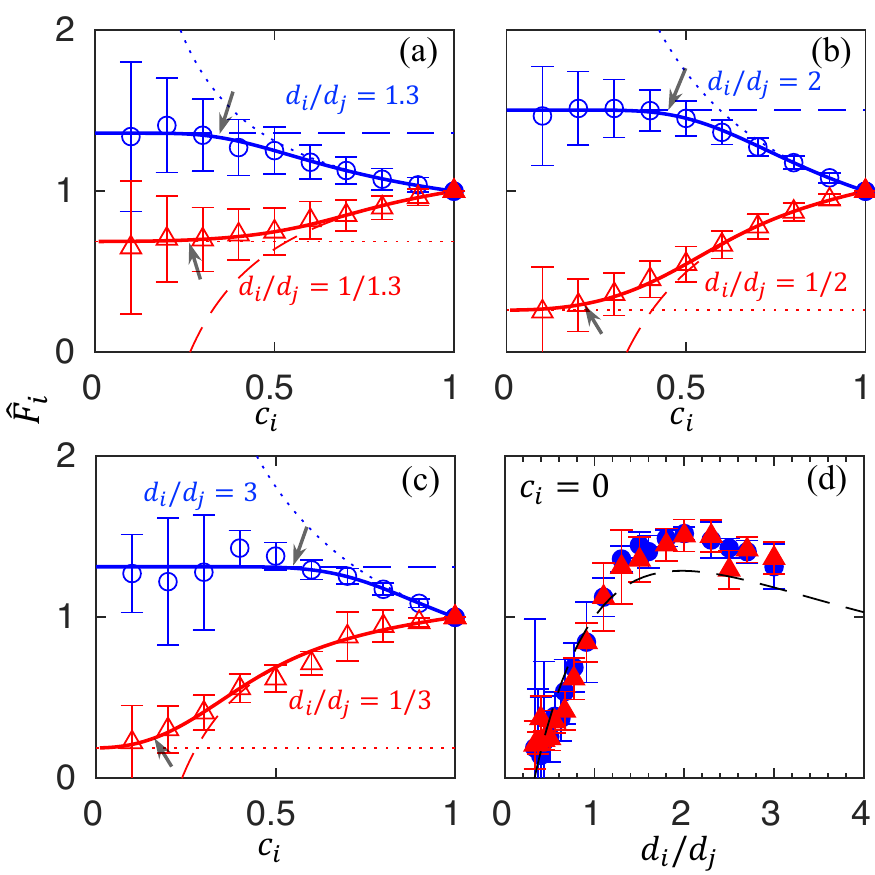}}
    \caption{ 
	(a-c) Scaled particle segregation force $\hat F_i=F_{\mathrm{seg},i}/ m_i g$ vs.\ species concentration $c_i$ for large (\textcolor{blue}{$\Circle$}) and small (\textcolor{red}{$\triangle$}) particles. Error bars are the standard deviation for the time-average of $F_{\mathrm{res},i}$. Dashed and dotted curves are predictions of the single intruder segregation force model extended to mixtures [Eq.~(\ref{balance})]. Solid curves are fits of Eq.~(\ref{tanh}) using large particle data. Arrows indicate the concentration $c_{i,\mathrm{crit}}$ where $F_i$ deviates from the intruder limit, see text. (d) $\hat F_{i,0}$ from fits of Eq.~(\ref{tanh}) to large (\textcolor{blue}{blue}) and small (\textcolor{red}{red}) particle data. Dashed curve is a single intruder model based on single intruder simulations~\cite{jing2020rising}.
    }
    \label{fig2}
\end{figure}

{\it Results}. The first key result of this paper is measurements of the dependence of the segregation force on concentration for various particle size ratios. Figure~\ref{fig2}(a-c) shows examples of $\hat F_{i}$ (symbols) vs.\ concentration for three size ratios ($d_l/d_s=1.3$, 2, and 3), where the error bars reflect fluctuations of the reactive restoring force in Fig.~\ref{scheme}(b). Although the error bars are largest at low concentrations, $\hat F_{i}$  clearly plateaus to a maximal (minimal) value approaching the single intruder limit $\hat F_{i,0}$ at $c_i\approx0$ and decreases (increases) monotonically with $c_i$ for large (small) particles. For both small and large species, $\hat F_{i,1}=1$ (or, equivalently, $F_{\mathrm{seg},i}=m_i g$) in the monodisperse limit ($c_i = 1$), since the segregation force exactly offsets the weight. 

Details of the dependence of $\hat F_{i}$ on $c_i$ vary with the size ratio, $d_i/d_j$. Specifically, Fig.~\ref{fig2}(a) for $d_l/d_s=1.3$ shows that the plateau in $\hat F_i$ for both species extends to a concentration of $c_{i}\approx0.3$. To quantify the extent of this plateau where particles of the lower concentration species are intruder-like (i.e., non-interacting with each other), we define $c_{i,\mathrm{crit}}$ as the critical concentration at which $\hat F_{i}-1$ deviates by 5\% from $\hat F_{i,0}-1$. For $c_i<c_{i,\mathrm{crit}}$, particles of species $i$ interact so infrequently with each other that the segregation force acting on them is essentially that for a single intruder particle.  As $c_i$ increases beyond $c_{i,\mathrm{crit}}$, interactions between particles of species $i$ become significant, eventually resulting in the segregation force approaching the monodisperse limit as $c_i$ approaches one.  This is the second key result:  there is a plateau in species concentration over which the segregation force for that species is simply that of a single intruder particle.  The plateau extends to higher concentrations as $d_i/d_j$ is increased, see Fig.~\ref{fig2}(b,c).  Furthermore, $c_{l,\mathrm{crit}}\ge c_{s,\mathrm{crit}}$, indicating that large particles act like intruders at higher concentrations than small particles.  For example, for $d_l/d_s=3$ [Fig.~\ref{fig2}(c)] the plateau for large particles extends to $c_{l,\mathrm{crit}}\approx 0.6$ nearly four times $c_{s,\mathrm{crit}}\approx 0.15$.

The total segregation force across both species for the entire system, which sums to the total particle weight, can be expressed 
using Eqs.~(\ref{eq1}) and (\ref{equilibrium}), as
\begin{equation}
\hat F_i c_i+ \hat F_jc_j  = 1.
\label{sumto1} 
\end{equation}
Noting that $c_j=1-c_i$ and $\hat F_{j}=F_{j,0}$ for $c_j\le c_{j,\mathrm{crit}}$ (or, equivalently, $c_i\ge 1-c_{j,\mathrm{crit}}$), we can predict $\hat F_{i}$ for mixtures not only in the intruder regime of species $i$, but also in the intruder regime of species $j$,
\begin{equation}
\hat F_{i}=\bigg\{
\begin{array}{lcl}
\hat F_{i,0} &  &  c_i\le c_{i,\mathrm{crit}}, \\
\big[1-\hat F_{j,0}(1-c_i)\big]/c_i &  &  c_i\ge 1-c_{j,\mathrm{crit}}.
\end{array}
\label{balance}
\end{equation}
Figures~\ref{fig2}(a-c) show that the predictions of Eq.~(\ref{balance}) for both large (dashed curves) and small particles (dotted curves) match the segregation force data (symbols) for large values of concentration when $\hat F_{i,0}$ and $\hat F_{j,0}$ are set to the intruder-limit values given in Fig.~\ref{fig2}(d) (described shortly). That is, selecting $\hat F_{l,0}$ at $c_i<c_{i,\mathrm{crit}}$ for large particles (dashed blue horizontal line) leads to the corresponding prediction for $\hat F_s$ at large $c_i$ (dashed red curve) and likewise for small particles (dotted red horizontal line and dotted blue curve).  This approximation fits the data well, except in the middle of the concentration range where the initial deviation of the data from the horizontal line reflects the approximate value of $c_{i,\mathrm{crit}}$.

Though Eq.~(\ref{balance}) combined with $\hat F_{i,0}$ and $\hat F_{j,0}$ predict $\hat F_i$ at the concentration extremes, a greater challenge is to model $\hat F_i$ in the intermediate transition regime (i.e., $c_{i,\mathrm{crit}} < c_i < 1-c_{j,\mathrm{crit}}$). Since $\hat F_i$ is bounded at both ends of the concentration range, we propose an empirical relation of the form
\begin{equation}
\hat F_{i}=\bigg\{
\begin{array}{lcl}
1+(\hat F_{i,0}-1) \tanh\bigg(\frac{1-\hat F_{j,0}}{\hat F_{i,0}-1} \frac{c_j}{c_i}\bigg), &  d_i/d_j\ge 1, \\
1 -(\hat F_{j,0}-1)  \tanh\bigg(\frac{1-\hat F_{i,0}}{\hat F_{j,0}-1}  \frac{c_i}{c_j}\bigg)\frac{c_j}{c_i}, &   d_i/d_j<1.
\end{array}
\label{tanh}
\end{equation}
Equation~(\ref{tanh}) applies to both large and small particles ($i=l$ for $d_i/d_j\ge 1$ and $i=s$ for $d_i/d_j<1$) and automatically satisfies the constraints that $\hat F_i=\hat F_{i,0}$ at $c_i=0$ and $\hat F_i=1$ at $c_i=1$. 
Model coefficients $\hat F_{i,0}$ and $\hat F_{j,0}$ correspond to intruder segregation forces and can be obtained by fitting Eq.~(\ref{tanh}) for $d_i/d_j\ge 1$ to the data for large particles or, equivalently, Eq.~(\ref{tanh}) for $d_i/d_j< 1$ to the data for small particles with no significant differences in the fit quality or fit parameters.

To demonstrate the validity of our simulation and fitting approach, Fig.~\ref{fig2}(d) shows $\hat F_{i,0}$ based on curve fits to Eq.~(\ref{tanh}) for both large (blue circles) and small (red triangles) particle data. The two data sets match within the uncertainty, demonstrating the robust nature of the hyperbolic functional form of Eq.~(\ref{tanh}) in characterizing the segregation force. In addition, the results match predictions (dashed curve) of a single intruder model  derived from single intruder simulations \cite{jing2020rising} with  particle properties (i.e., $d_l=1-40\,$mm, $d_s=5\,$mm, and $\rho=2.5\,$g/mm$^3$), contact model (i.e., Hertz contact model with Young's modulus of 5$\times10^7\,$Pa and Poison's ratio 0.4), and flow geometry (inclined chute) different from this study.
This validates not only the values we find for the segregation force at the single intruder limit, but also our approach for direct measurement of segregation forces in bidisperse mixtures.

\begin{figure}
    \centerline{\includegraphics[width=3.4 in]{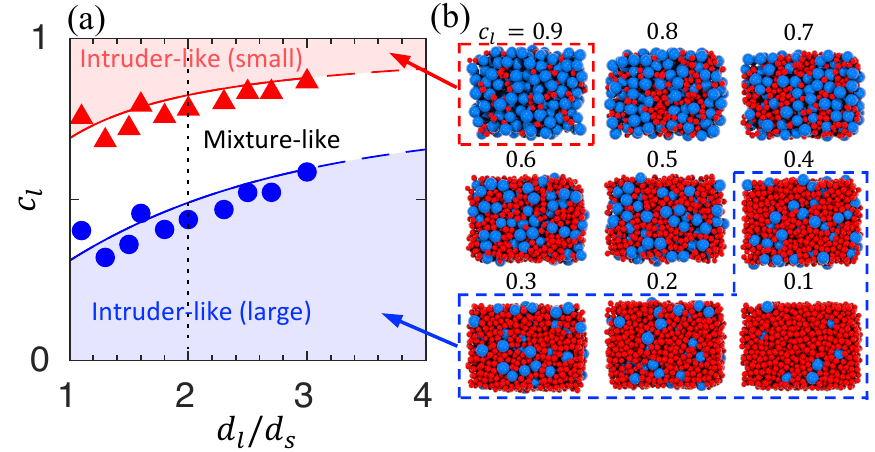}}
    \caption{
    (a) Phase diagram showing segregation force regimes (shaded areas) dependence on large particle concentration and size ratio. Symbols represent $c_{i,\mathrm{crit}}$ for large (\textcolor{blue}{blue}) and small (\textcolor{red}{red}) particles. Curves (from fits) are $c_{i,\mathrm{crit}}=0.74[1-\exp(-0.52 d_i/d_j)]$.
    (b) Sheared bed images for $d_l/d_s=2$ [vertical dotted line in (a)] at $c_l$ intervals of 0.1. For $c_s<c_{s,\mathrm{crit}}\approx 0.18$ (or, equivalently, $c_l\gtrapprox0.82$) the small particle (\textcolor{red}{red}) segregation force equals that on a single small intruder, while for $c_l<c_{l,\mathrm{crit}}\approx 0.46$ the large particle segregation force equals that on a single large intruder.  Intermediate concentrations ($0.46\lessapprox c_l\lessapprox0.82$), where segregation forces are less than for intruders, are termed mixture-like. 
    }
    \label{figure3}
\end{figure}

We determine the critical concentration, $c_{i,\mathrm{crit}}$ below which the segregation force is nearly independent of concentration by conducting 260 simulations at different concentrations, size ratios, and shear rates, and fitting the resulting segregation force data to Eq.~(\ref{tanh}).  
The phase diagram in Fig.~\ref{figure3}(a) shows that $c_{i,\mathrm{crit}}$ (symbols) for both large and small particles increases monotonically with size ratio for the range explored here ($1< d_l/d_s<3$) and is reasonably well fit by the expression $c_{i,\mathrm{crit}}=0.74[1-\exp(-0.54 d_i/d_j)].$  The limiting value of $c_{i,\mathrm{crit}}=0.74$ for $d_i/d_j\gg1$ matches the free sifting limit for small particles in a network of randomly close-packed large particles at $\phi_\mathrm{max}=0.64$, i.e., $1/(2-\phi_\mathrm{max})$ \cite{prasad2017subjamming}. This suggests that for $c_l>0.74$ small particles percolate downward through the voids without significantly affecting the flow of large particles, indicating a possible change in the size segregation mechanism~\cite{golick2009mixing,schlick2015modeling}.

\begin{figure}
    \centerline{\includegraphics[width=3.4 in]{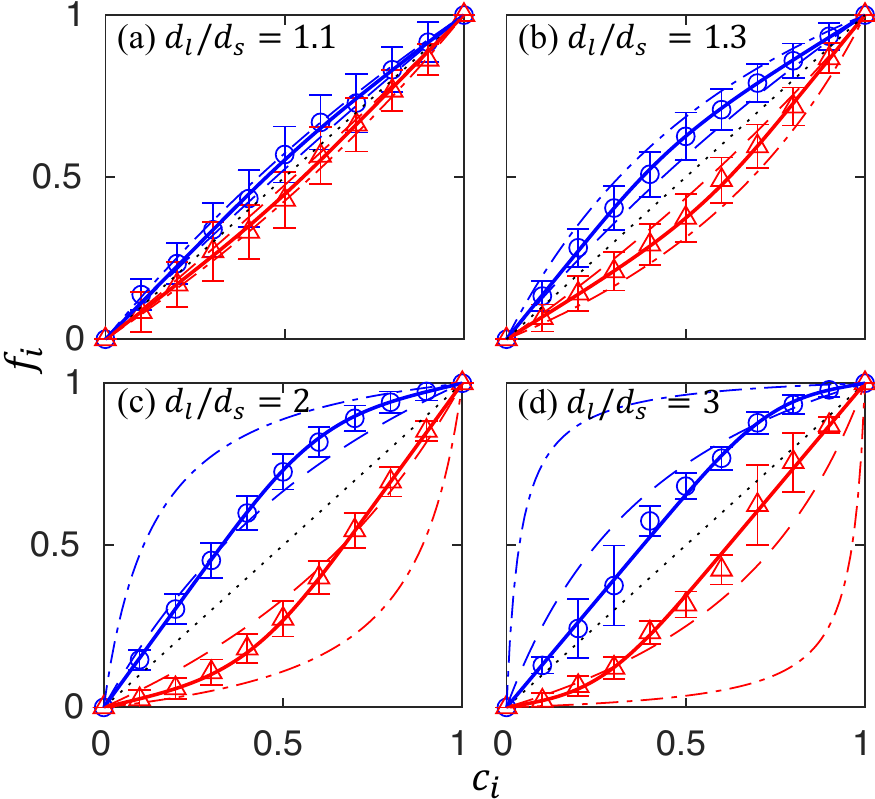}}
    \caption{Ratio of species specific pressure to bulk pressure, $f_i=P_i/P$, for different size ratios. Symbols represent data for large (\textcolor{blue}{blue}) and small (\textcolor{red}{red}) particles. 
    Solid curves are predictions of Eq.~(\ref{tanh}) recast as a pressure ratio, i.e., $f_i=c_i \hat F_i$.
    Thin curves are $f_i$ as assumed in previous studies: $f_i=d_ic_i/\sum d_i c_i$  (dashed) \cite{marks2012grainsize} and $f_i=d_i^3c_i/\sum d_i^3 c_i$  (dash-dot) \cite{tunuguntla2014mixture}.
     Dotted line represents the monodisperse case ($d_l/d_s=1$) where $f_i=c_i$.	
    }
    \label{fig4}
\end{figure}

Alternately, in the monodisperse mixture limit ($d_i/d_j=1$), the exponential fit to the data in Fig.~\ref{figure3}(a) gives $c_{i,\mathrm{crit}}\approx0.31$, matching the  concentration at which mixtures of equal diameter conducting and insulating spheres become globally conductive (exceed the percolation threshold)~\cite{powell1979site,ziff2017percolation}. Further, the critical concentrations for $1/3\le d_s/d_l< 1$ from this study also match the percolation thresholds in size-bidisperse mixtures~\cite{he2004effect}, suggesting that the particle segregation force and geometric percolation are related. Anecdotal support for this picture is provided by Fig.~\ref{figure3}(b), which shows shear flow images for $d_l/d_s=2$.  In the intruder-like regime for small particles (large $c_l$), small particles appear to contact each other only in the voids between large particles, whereas in the intruder-like regime for large particles (small $c_l$) large particles are well-separated by a continuous phase of small particles and are unlikely to interact directly with each other.  Further investigation of the connection between the intruder regimes and the percolation limit is clearly necessary but is beyond the scope of this Letter.

{\it Discussion}.  Our results characterizing the segregation force can be applied to continuum descriptions of segregation.  Previous studies assume $\hat F_{i}$ depends linearly on $c_i$ to close the momentum equation~\cite{gray2005theory,rousseau2021bridging}. Despite some success for these continuum models in predicting concentration profiles of equal-volume mixtures, the assumed linear relation between $\hat F_i$ and $c_i$ fails to capture the segregation force plateau for intruders. In addition, the resulting symmetric form for the species-specific pressure, when coupled with a linear drag model, fails to predict the asymmetric concentration dependence of segregation (i.e., small particles among mostly large particles segregate faster than {\it vice versa})~\cite{golick2009mixing,gajjar2014asymmetric,jones2018asymmetric}. To address the asymmetric segregation flux, $\hat F_{i}$ was later proposed to be quadratic in $c_i$ \cite{gajjar2014asymmetric,tripathi2021theory,duan2021modelling}. 
Though the coefficients in a quadratic model can be adjusted to minimize the difference between the model and the data in the plateau regime, the model is never truly independent of $c_i$, as must be true approaching the intruder limit ($c_i\approx0$).

To address these shortcomings in modeling the segregation force within the framework of the continuum model, we recast our results (data and model [Eq.~(\ref{tanh})]) as partial pressures (normal stresses), i.e., ${\partial P_i}/{\partial z}=N_iF_{\mathrm{seg},i}/LWH={n_iF_{\mathrm{seg},i}}$ \cite{rousseau2021bridging}, where  $n_i=c_i\phi /V_i$ is the particle number density.  Combined with the bulk pressure gradient $\partial P/ \partial z= \phi \rho g $, the ratio of the pressure contribution of species $i$ to the bulk pressure, or normal stress fraction, is $f_i=P_i/P=c_i \hat F_{i}$ \cite{tunuguntla2017comparing}, which, unlike the standard mixture theory, does not necessarily equal the species volume fraction.  

Here, having measured $\hat F_i$ vs.\ $c_i$, we can directly evaluate $f_i$ as Fig.~\ref{fig4} shows for four examples at $d_l/d_s=1.1$, 1.3, 2, and 3. At all concentrations, the pressure partition functions for large and small particles sum to 1 (i.e., $ f_l+f_s=1$), and the curves based on the segregation force model of Eq.~(\ref{tanh}) match the simulation data -- the third key result of this paper.  The deviation of the pressure partitioning for $d_l/d_s \ne 1$ from the linear monodisperse case, $f_i=c_i$ (dotted line) and from previously proposed models assuming $f_i$ is a weighted function of particle size, $f_i=d_ic_i/\sum d_i c_i$ (dashed) \cite{marks2012grainsize}, or volume, $f_i=d_i^3c_i/\sum d_i^3 c_i$ (dash-dot) \cite{tunuguntla2014mixture}, is clear for all size ratios. Thus, the more accurate pressure partition function based on Eq.~(\ref{tanh}) can be directly applied to continuum models of flowing mixtures of bidisperse granular materials such as those used for free surface flows \cite{marks2012grainsize,tunuguntla2014mixture,tripathi2021theory,rousseau2021bridging}.

Our results capture and characterize the concentration dependence of the segregation force in uniform shear flows, but a word of caution about extensions is in order. Recent studies indicate that the intruder segregation force $\hat F_{i,0}$ also depends on the shear rate gradient~\cite{fan2011theory,guillard2016scaling,jing2021unified}. Although the shear rate gradient-induced component of $F_\mathrm{seg}$ is negligible in most free surface flows~\cite{jing2020rising}, further study of the concentration dependence in flows with shear rate gradients is clearly necessary. 

\textbf{Acknowledgements.}
Supported by the National Science Foundation under Grant No.~CBET-1929265.

\nocite{*}

\bibliography{apssamp}

\providecommand{\noopsort}[1]{}\providecommand{\singleletter}[1]{#1}%
\begin{thebibliography}{38}%
\makeatletter
\providecommand \@ifxundefined [1]{%
 \@ifx{#1\undefined}
}%
\providecommand \@ifnum [1]{%
 \ifnum #1\expandafter \@firstoftwo
 \else \expandafter \@secondoftwo
 \fi
}%
\providecommand \@ifx [1]{%
 \ifx #1\expandafter \@firstoftwo
 \else \expandafter \@secondoftwo
 \fi
}%
\providecommand \natexlab [1]{#1}%
\providecommand \enquote  [1]{``#1''}%
\providecommand \bibnamefont  [1]{#1}%
\providecommand \bibfnamefont [1]{#1}%
\providecommand \citenamefont [1]{#1}%
\providecommand \href@noop [0]{\@secondoftwo}%
\providecommand \href [0]{\begingroup \@sanitize@url \@href}%
\providecommand \@href[1]{\@@startlink{#1}\@@href}%
\providecommand \@@href[1]{\endgroup#1\@@endlink}%
\providecommand \@sanitize@url [0]{\catcode `\\12\catcode `\$12\catcode
  `\&12\catcode `\#12\catcode `\^12\catcode `\_12\catcode `\%12\relax}%
\providecommand \@@startlink[1]{}%
\providecommand \@@endlink[0]{}%
\providecommand \url  [0]{\begingroup\@sanitize@url \@url }%
\providecommand \@url [1]{\endgroup\@href {#1}{\urlprefix }}%
\providecommand \urlprefix  [0]{URL }%
\providecommand \Eprint [0]{\href }%
\providecommand \doibase [0]{https://doi.org/}%
\providecommand \selectlanguage [0]{\@gobble}%
\providecommand \bibinfo  [0]{\@secondoftwo}%
\providecommand \bibfield  [0]{\@secondoftwo}%
\providecommand \translation [1]{[#1]}%
\providecommand \BibitemOpen [0]{}%
\providecommand \bibitemStop [0]{}%
\providecommand \bibitemNoStop [0]{.\EOS\space}%
\providecommand \EOS [0]{\spacefactor3000\relax}%
\providecommand \BibitemShut  [1]{\csname bibitem#1\endcsname}%
\let\auto@bib@innerbib\@empty
\bibitem [{\citenamefont {Ottino}\ and\ \citenamefont
  {Khakhar}(2000)}]{ottino2000mixing}%
  \BibitemOpen
  \bibfield  {author} {\bibinfo {author} {\bibfnamefont {J.~M.}\ \bibnamefont
  {Ottino}}\ and\ \bibinfo {author} {\bibfnamefont {D.~V.}\ \bibnamefont
  {Khakhar}},\ }\bibinfo {title} {Mixing and segregation of granular
  materials},\ \href@noop {} {\bibfield  {journal} {\bibinfo  {journal} {Annu.
  Rev. Fluid Mech.}\ }\textbf {\bibinfo {volume} {32}},\ \bibinfo {pages} {55}
  (\bibinfo {year} {2000})}\BibitemShut {NoStop}%
\bibitem [{\citenamefont {Ottino}\ and\ \citenamefont
  {Lueptow}(2008)}]{ottino2008mixing}%
  \BibitemOpen
  \bibfield  {author} {\bibinfo {author} {\bibfnamefont {J.~M.}\ \bibnamefont
  {Ottino}}\ and\ \bibinfo {author} {\bibfnamefont {R.~M.}\ \bibnamefont
  {Lueptow}},\ }\bibinfo {title} {On mixing and demixing},\ \href@noop {}
  {\bibfield  {journal} {\bibinfo  {journal} {Science}\ }\textbf {\bibinfo
  {volume} {319}},\ \bibinfo {pages} {912} (\bibinfo {year}
  {2008})}\BibitemShut {NoStop}%
\bibitem [{\citenamefont {Frey}\ and\ \citenamefont
  {Church}(2009)}]{frey2009river}%
  \BibitemOpen
  \bibfield  {author} {\bibinfo {author} {\bibfnamefont {P.}~\bibnamefont
  {Frey}}\ and\ \bibinfo {author} {\bibfnamefont {M.}~\bibnamefont {Church}},\
  }\bibinfo {title} {How river beds move},\ \href@noop {} {\bibfield  {journal}
  {\bibinfo  {journal} {Science}\ }\textbf {\bibinfo {volume} {325}},\ \bibinfo
  {pages} {1509} (\bibinfo {year} {2009})}\BibitemShut {NoStop}%
\bibitem [{\citenamefont {Gray}(2018)}]{gray2018particle}%
  \BibitemOpen
  \bibfield  {author} {\bibinfo {author} {\bibfnamefont {J.~M. N.~T.}\
  \bibnamefont {Gray}},\ }\bibinfo {title} {Particle segregation in dense
  granular flows},\ \href@noop {} {\bibfield  {journal} {\bibinfo  {journal}
  {Annu. Rev. Fluid Mech.}\ }\textbf {\bibinfo {volume} {50}},\ \bibinfo
  {pages} {407} (\bibinfo {year} {2018})}\BibitemShut {NoStop}%
\bibitem [{\citenamefont {Umbanhowar}\ \emph {et~al.}(2019)\citenamefont
  {Umbanhowar}, \citenamefont {Lueptow},\ and\ \citenamefont
  {Ottino}}]{umbanhowar2019modeling}%
  \BibitemOpen
  \bibfield  {author} {\bibinfo {author} {\bibfnamefont {P.~B.}\ \bibnamefont
  {Umbanhowar}}, \bibinfo {author} {\bibfnamefont {R.~M.}\ \bibnamefont
  {Lueptow}},\ and\ \bibinfo {author} {\bibfnamefont {J.~M.}\ \bibnamefont
  {Ottino}},\ }\bibinfo {title} {Modeling segregation in granular flows},\
  \href@noop {} {\bibfield  {journal} {\bibinfo  {journal} {Annu. Rev. Chem.
  Biomol. Eng.}\ }\textbf {\bibinfo {volume} {10}},\ \bibinfo {pages} {129}
  (\bibinfo {year} {2019})}\BibitemShut {NoStop}%
\bibitem [{\citenamefont {Thornton}(2021)}]{thornton2021brief}%
  \BibitemOpen
  \bibfield  {author} {\bibinfo {author} {\bibfnamefont {A.}~\bibnamefont
  {Thornton}},\ }\bibinfo {title} {A brief review of (multi-scale) modelling
  approaches to segregation},\ \href@noop {} {\bibfield  {journal} {\bibinfo
  {journal} {EPJ Web Conf.}\ }\textbf {\bibinfo {volume} {249}},\ \bibinfo
  {pages} {01004} (\bibinfo {year} {2021})}\BibitemShut {NoStop}%
\bibitem [{\citenamefont {Tripathi}\ and\ \citenamefont
  {Khakhar}(2011)}]{tripathi2011numerical}%
  \BibitemOpen
  \bibfield  {author} {\bibinfo {author} {\bibfnamefont {A.}~\bibnamefont
  {Tripathi}}\ and\ \bibinfo {author} {\bibfnamefont {D.~V.}\ \bibnamefont
  {Khakhar}},\ }\bibinfo {title} {Numerical simulation of the sedimentation of
  a sphere in a sheared granular fluid: a granular stokes experiment},\
  \href@noop {} {\bibfield  {journal} {\bibinfo  {journal} {Phys. Rev. Lett.}\
  }\textbf {\bibinfo {volume} {107}},\ \bibinfo {pages} {108001} (\bibinfo
  {year} {2011})}\BibitemShut {NoStop}%
\bibitem [{\citenamefont {Guillard}\ \emph {et~al.}(2016)\citenamefont
  {Guillard}, \citenamefont {Forterre},\ and\ \citenamefont
  {Pouliquen}}]{guillard2016scaling}%
  \BibitemOpen
  \bibfield  {author} {\bibinfo {author} {\bibfnamefont {F.}~\bibnamefont
  {Guillard}}, \bibinfo {author} {\bibfnamefont {Y.}~\bibnamefont {Forterre}},\
  and\ \bibinfo {author} {\bibfnamefont {O.}~\bibnamefont {Pouliquen}},\
  }\bibinfo {title} {Scaling laws for segregation forces in dense sheared
  granular flows},\ \href@noop {} {\bibfield  {journal} {\bibinfo  {journal}
  {J. Fluid Mech.}\ }\textbf {\bibinfo {volume} {807}},\ \bibinfo {pages} {R1}
  (\bibinfo {year} {2016})}\BibitemShut {NoStop}%
\bibitem [{\citenamefont {Jing}\ \emph {et~al.}(2017)\citenamefont {Jing},
  \citenamefont {Kwok},\ and\ \citenamefont {Leung}}]{jing2017micromechanical}%
  \BibitemOpen
  \bibfield  {author} {\bibinfo {author} {\bibfnamefont {L.}~\bibnamefont
  {Jing}}, \bibinfo {author} {\bibfnamefont {C.~Y.}\ \bibnamefont {Kwok}},\
  and\ \bibinfo {author} {\bibfnamefont {Y.~F.}\ \bibnamefont {Leung}},\
  }\bibinfo {title} {Micromechanical origin of particle size segregation},\
  \href@noop {} {\bibfield  {journal} {\bibinfo  {journal} {Phys. Rev. Lett.}\
  }\textbf {\bibinfo {volume} {118}},\ \bibinfo {pages} {118001} (\bibinfo
  {year} {2017})}\BibitemShut {NoStop}%
\bibitem [{\citenamefont {van~der Vaart}\ \emph {et~al.}(2018)\citenamefont
  {van~der Vaart}, \citenamefont {van Schrojenstein~Lantman}, \citenamefont
  {Weinhart}, \citenamefont {Luding}, \citenamefont {Ancey},\ and\
  \citenamefont {Thornton}}]{van2018segregation}%
  \BibitemOpen
  \bibfield  {author} {\bibinfo {author} {\bibfnamefont {K.}~\bibnamefont
  {van~der Vaart}}, \bibinfo {author} {\bibfnamefont {M.~P.}\ \bibnamefont {van
  Schrojenstein~Lantman}}, \bibinfo {author} {\bibfnamefont {T.}~\bibnamefont
  {Weinhart}}, \bibinfo {author} {\bibfnamefont {S.}~\bibnamefont {Luding}},
  \bibinfo {author} {\bibfnamefont {C.}~\bibnamefont {Ancey}},\ and\ \bibinfo
  {author} {\bibfnamefont {A.~R.}\ \bibnamefont {Thornton}},\ }\bibinfo {title}
  {Segregation of large particles in dense granular flows suggests a granular
  saffman effect},\ \href@noop {} {\bibfield  {journal} {\bibinfo  {journal}
  {Phys. Rev. Fluids}\ }\textbf {\bibinfo {volume} {3}},\ \bibinfo {pages}
  {074303} (\bibinfo {year} {2018})}\BibitemShut {NoStop}%
\bibitem [{\citenamefont {Staron}(2018)}]{staron2018rising}%
  \BibitemOpen
  \bibfield  {author} {\bibinfo {author} {\bibfnamefont {L.}~\bibnamefont
  {Staron}},\ }\bibinfo {title} {Rising dynamics and lift effect in dense
  segregating granular flows},\ \href@noop {} {\bibfield  {journal} {\bibinfo
  {journal} {Phys. Fluids}\ }\textbf {\bibinfo {volume} {30}},\ \bibinfo
  {pages} {123303} (\bibinfo {year} {2018})}\BibitemShut {NoStop}%
\bibitem [{\citenamefont {Jing}\ \emph {et~al.}(2020)\citenamefont {Jing},
  \citenamefont {Ottino}, \citenamefont {Lueptow},\ and\ \citenamefont
  {Umbanhowar}}]{jing2020rising}%
  \BibitemOpen
  \bibfield  {author} {\bibinfo {author} {\bibfnamefont {L.}~\bibnamefont
  {Jing}}, \bibinfo {author} {\bibfnamefont {J.~M.}\ \bibnamefont {Ottino}},
  \bibinfo {author} {\bibfnamefont {R.~M.}\ \bibnamefont {Lueptow}},\ and\
  \bibinfo {author} {\bibfnamefont {P.~B.}\ \bibnamefont {Umbanhowar}},\
  }\bibinfo {title} {Rising and sinking intruders in dense granular flows},\
  \href@noop {} {\bibfield  {journal} {\bibinfo  {journal} {Phys. Rev. Res.}\
  }\textbf {\bibinfo {volume} {2}},\ \bibinfo {pages} {022069} (\bibinfo {year}
  {2020})}\BibitemShut {NoStop}%
\bibitem [{\citenamefont {Jing}\ \emph {et~al.}(2021)\citenamefont {Jing},
  \citenamefont {Ottino}, \citenamefont {Lueptow},\ and\ \citenamefont
  {Umbanhowar}}]{jing2021unified}%
  \BibitemOpen
  \bibfield  {author} {\bibinfo {author} {\bibfnamefont {L.}~\bibnamefont
  {Jing}}, \bibinfo {author} {\bibfnamefont {J.~M.}\ \bibnamefont {Ottino}},
  \bibinfo {author} {\bibfnamefont {R.~M.}\ \bibnamefont {Lueptow}},\ and\
  \bibinfo {author} {\bibfnamefont {P.~B.}\ \bibnamefont {Umbanhowar}},\
  }\bibinfo {title} {A unified description of gravity-and kinematics-induced
  segregation forces in dense granular flows},\ \href@noop {} {\bibfield
  {journal} {\bibinfo  {journal} {J. Fluid Mech.}\ } (\bibinfo {year}
  {2021})},\ \bibinfo {note} {in press}\BibitemShut {NoStop}%
\bibitem [{\citenamefont {Tripathi}\ \emph {et~al.}(2021)\citenamefont
  {Tripathi}, \citenamefont {Kumar}, \citenamefont {Nema},\ and\ \citenamefont
  {Khakhar}}]{tripathi2021theory}%
  \BibitemOpen
  \bibfield  {author} {\bibinfo {author} {\bibfnamefont {A.}~\bibnamefont
  {Tripathi}}, \bibinfo {author} {\bibfnamefont {A.}~\bibnamefont {Kumar}},
  \bibinfo {author} {\bibfnamefont {M.}~\bibnamefont {Nema}},\ and\ \bibinfo
  {author} {\bibfnamefont {D.}~\bibnamefont {Khakhar}},\ }\bibinfo {title}
  {Theory for size segregation in flowing granular mixtures based on
  computation of forces on a single large particle},\ \href@noop {} {\bibfield
  {journal} {\bibinfo  {journal} {Phys. Res. E}\ }\textbf {\bibinfo {volume}
  {103}},\ \bibinfo {pages} {L031301} (\bibinfo {year} {2021})}\BibitemShut
  {NoStop}%
\bibitem [{\citenamefont {Rousseau}\ \emph {et~al.}(2021)\citenamefont
  {Rousseau}, \citenamefont {Chassagne}, \citenamefont {Chauchat},
  \citenamefont {Maurin},\ and\ \citenamefont {Frey}}]{rousseau2021bridging}%
  \BibitemOpen
  \bibfield  {author} {\bibinfo {author} {\bibfnamefont {H.}~\bibnamefont
  {Rousseau}}, \bibinfo {author} {\bibfnamefont {R.}~\bibnamefont {Chassagne}},
  \bibinfo {author} {\bibfnamefont {J.}~\bibnamefont {Chauchat}}, \bibinfo
  {author} {\bibfnamefont {R.}~\bibnamefont {Maurin}},\ and\ \bibinfo {author}
  {\bibfnamefont {P.}~\bibnamefont {Frey}},\ }\bibinfo {title} {Bridging the
  gap between particle-scale forces and continuum modelling of size
  segregation: application to bedload transport},\ \href@noop {} {\bibfield
  {journal} {\bibinfo  {journal} {J. Fluid Mech.}\ }\textbf {\bibinfo {volume}
  {916}},\ \bibinfo {pages} {A26} (\bibinfo {year} {2021})}\BibitemShut
  {NoStop}%
\bibitem [{\citenamefont {van~der Vaart}\ \emph {et~al.}(2015)\citenamefont
  {van~der Vaart}, \citenamefont {Gajjar}, \citenamefont {Epely-Chauvin},
  \citenamefont {Andreini}, \citenamefont {Gray},\ and\ \citenamefont
  {Ancey}}]{van2015underlying}%
  \BibitemOpen
  \bibfield  {author} {\bibinfo {author} {\bibfnamefont {K.}~\bibnamefont
  {van~der Vaart}}, \bibinfo {author} {\bibfnamefont {P.}~\bibnamefont
  {Gajjar}}, \bibinfo {author} {\bibfnamefont {G.}~\bibnamefont
  {Epely-Chauvin}}, \bibinfo {author} {\bibfnamefont {N.}~\bibnamefont
  {Andreini}}, \bibinfo {author} {\bibfnamefont {J.~M. N.~T.}\ \bibnamefont
  {Gray}},\ and\ \bibinfo {author} {\bibfnamefont {C.}~\bibnamefont {Ancey}},\
  }\bibinfo {title} {Underlying asymmetry within particle size segregation},\
  \href@noop {} {\bibfield  {journal} {\bibinfo  {journal} {Phys. Rev. Lett.}\
  }\textbf {\bibinfo {volume} {114}},\ \bibinfo {pages} {238001} (\bibinfo
  {year} {2015})}\BibitemShut {NoStop}%
\bibitem [{\citenamefont {Jones}\ \emph {et~al.}(2018)\citenamefont {Jones},
  \citenamefont {Isner}, \citenamefont {Xiao}, \citenamefont {Ottino},
  \citenamefont {Umbanhowar},\ and\ \citenamefont
  {Lueptow}}]{jones2018asymmetric}%
  \BibitemOpen
  \bibfield  {author} {\bibinfo {author} {\bibfnamefont {R.~P.}\ \bibnamefont
  {Jones}}, \bibinfo {author} {\bibfnamefont {A.~B.}\ \bibnamefont {Isner}},
  \bibinfo {author} {\bibfnamefont {H.}~\bibnamefont {Xiao}}, \bibinfo {author}
  {\bibfnamefont {J.~M.}\ \bibnamefont {Ottino}}, \bibinfo {author}
  {\bibfnamefont {P.~B.}\ \bibnamefont {Umbanhowar}},\ and\ \bibinfo {author}
  {\bibfnamefont {R.~M.}\ \bibnamefont {Lueptow}},\ }\bibinfo {title}
  {Asymmetric concentration dependence of segregation fluxes in granular
  flows},\ \href {https://doi.org/10.1103/PhysRevFluids.3.094304} {\bibfield
  {journal} {\bibinfo  {journal} {Phys. Rev. Fluids}\ }\textbf {\bibinfo
  {volume} {3}},\ \bibinfo {pages} {094304} (\bibinfo {year}
  {2018})}\BibitemShut {NoStop}%
\bibitem [{\citenamefont {Isner}\ \emph {et~al.}(2020)\citenamefont {Isner},
  \citenamefont {Umbanhowar}, \citenamefont {Ottino},\ and\ \citenamefont
  {Lueptow}}]{isner2020axisymmetric}%
  \BibitemOpen
  \bibfield  {author} {\bibinfo {author} {\bibfnamefont {A.~B.}\ \bibnamefont
  {Isner}}, \bibinfo {author} {\bibfnamefont {P.~B.}\ \bibnamefont
  {Umbanhowar}}, \bibinfo {author} {\bibfnamefont {J.~M.}\ \bibnamefont
  {Ottino}},\ and\ \bibinfo {author} {\bibfnamefont {R.~M.}\ \bibnamefont
  {Lueptow}},\ }\bibinfo {title} {Axisymmetric granular flow on a bounded
  conical heap: Kinematics and size segregation},\ \href@noop {} {\bibfield
  {journal} {\bibinfo  {journal} {Chem. Eng. Sci.}\ }\textbf {\bibinfo {volume}
  {217}},\ \bibinfo {pages} {115505} (\bibinfo {year} {2020})}\BibitemShut
  {NoStop}%
\bibitem [{\citenamefont {Cundall}\ and\ \citenamefont
  {Strack}(1979)}]{cundall1979discrete}%
  \BibitemOpen
  \bibfield  {author} {\bibinfo {author} {\bibfnamefont {P.~A.}\ \bibnamefont
  {Cundall}}\ and\ \bibinfo {author} {\bibfnamefont {O.~D.~L.}\ \bibnamefont
  {Strack}},\ }\bibinfo {title} {A discrete numerical model for granular
  assemblies},\ \href@noop {} {\bibfield  {journal} {\bibinfo  {journal}
  {Géotechnique}\ }\textbf {\bibinfo {volume} {29}},\ \bibinfo {pages} {47}
  (\bibinfo {year} {1979})}\BibitemShut {NoStop}%
\bibitem [{\citenamefont {Staron}\ and\ \citenamefont
  {Phillips}(2014)}]{staron2014segregation}%
  \BibitemOpen
  \bibfield  {author} {\bibinfo {author} {\bibfnamefont {L.}~\bibnamefont
  {Staron}}\ and\ \bibinfo {author} {\bibfnamefont {J.~C.}\ \bibnamefont
  {Phillips}},\ }\bibinfo {title} {Segregation time-scale in bi-disperse
  granular flows},\ \href@noop {} {\bibfield  {journal} {\bibinfo  {journal}
  {Phys. Fluids}\ }\textbf {\bibinfo {volume} {26}},\ \bibinfo {pages} {033302}
  (\bibinfo {year} {2014})}\BibitemShut {NoStop}%
\bibitem [{\citenamefont {Lerner}\ \emph {et~al.}(2012)\citenamefont {Lerner},
  \citenamefont {D{\"u}ring},\ and\ \citenamefont {Wyart}}]{Lerner2012unified}%
  \BibitemOpen
  \bibfield  {author} {\bibinfo {author} {\bibfnamefont {E.}~\bibnamefont
  {Lerner}}, \bibinfo {author} {\bibfnamefont {G.}~\bibnamefont {D{\"u}ring}},\
  and\ \bibinfo {author} {\bibfnamefont {M.}~\bibnamefont {Wyart}},\ }\bibinfo
  {title} {A unified framework for non-brownian suspension flows and soft
  amorphous solids},\ \href {https://doi.org/10.1073/pnas.1120215109}
  {\bibfield  {journal} {\bibinfo  {journal} {Proc. Natl. Acad. Sci.}\ }\textbf
  {\bibinfo {volume} {109}},\ \bibinfo {pages} {4798} (\bibinfo {year}
  {2012})}\BibitemShut {NoStop}%
\bibitem [{\citenamefont {Clark}\ \emph {et~al.}(2018)\citenamefont {Clark},
  \citenamefont {Thompson}, \citenamefont {Shattuck}, \citenamefont
  {Ouellette},\ and\ \citenamefont {O'Hern}}]{clark2018critical}%
  \BibitemOpen
  \bibfield  {author} {\bibinfo {author} {\bibfnamefont {A.~H.}\ \bibnamefont
  {Clark}}, \bibinfo {author} {\bibfnamefont {J.~D.}\ \bibnamefont {Thompson}},
  \bibinfo {author} {\bibfnamefont {M.~D.}\ \bibnamefont {Shattuck}}, \bibinfo
  {author} {\bibfnamefont {N.~T.}\ \bibnamefont {Ouellette}},\ and\ \bibinfo
  {author} {\bibfnamefont {C.~S.}\ \bibnamefont {O'Hern}},\ }\bibinfo {title}
  {Critical scaling near the yielding transition in granular media},\
  \href@noop {} {\bibfield  {journal} {\bibinfo  {journal} {Phys. Rev. E}\
  }\textbf {\bibinfo {volume} {97}},\ \bibinfo {pages} {062901} (\bibinfo
  {year} {2018})}\BibitemShut {NoStop}%
\bibitem [{\citenamefont {Fry}\ \emph {et~al.}(2018)\citenamefont {Fry},
  \citenamefont {Umbanhowar}, \citenamefont {Ottino},\ and\ \citenamefont
  {Lueptow}}]{fry2018effect}%
  \BibitemOpen
  \bibfield  {author} {\bibinfo {author} {\bibfnamefont {A.~M.}\ \bibnamefont
  {Fry}}, \bibinfo {author} {\bibfnamefont {P.~B.}\ \bibnamefont {Umbanhowar}},
  \bibinfo {author} {\bibfnamefont {J.~M.}\ \bibnamefont {Ottino}},\ and\
  \bibinfo {author} {\bibfnamefont {R.~M.}\ \bibnamefont {Lueptow}},\ }\bibinfo
  {title} {Effect of pressure on segregation in granular shear flows},\ \href
  {https://doi.org/10.1103/PhysRevE.97.062906} {\bibfield  {journal} {\bibinfo
  {journal} {Phys. Rev. E}\ }\textbf {\bibinfo {volume} {97}},\ \bibinfo
  {pages} {062906} (\bibinfo {year} {2018})}\BibitemShut {NoStop}%
\bibitem [{\citenamefont {Saitoh}\ and\ \citenamefont
  {Tighe}(2019)}]{saitoh2019nonlocal}%
  \BibitemOpen
  \bibfield  {author} {\bibinfo {author} {\bibfnamefont {K.}~\bibnamefont
  {Saitoh}}\ and\ \bibinfo {author} {\bibfnamefont {B.~P.}\ \bibnamefont
  {Tighe}},\ }\bibinfo {title} {Nonlocal effects in inhomogeneous flows of soft
  athermal disks},\ \href@noop {} {\bibfield  {journal} {\bibinfo  {journal}
  {Phys. Rev. Lett.}\ }\textbf {\bibinfo {volume} {122}},\ \bibinfo {pages}
  {188001} (\bibinfo {year} {2019})}\BibitemShut {NoStop}%
\bibitem [{\citenamefont {Duan}\ \emph {et~al.}(2020)\citenamefont {Duan},
  \citenamefont {Umbanhowar}, \citenamefont {Ottino},\ and\ \citenamefont
  {Lueptow}}]{duan2020segregation}%
  \BibitemOpen
  \bibfield  {author} {\bibinfo {author} {\bibfnamefont {Y.}~\bibnamefont
  {Duan}}, \bibinfo {author} {\bibfnamefont {P.~B.}\ \bibnamefont
  {Umbanhowar}}, \bibinfo {author} {\bibfnamefont {J.~M.}\ \bibnamefont
  {Ottino}},\ and\ \bibinfo {author} {\bibfnamefont {R.~M.}\ \bibnamefont
  {Lueptow}},\ }\bibinfo {title} {Segregation models for density-bidisperse
  granular flows},\ \href@noop {} {\bibfield  {journal} {\bibinfo  {journal}
  {Phys. Rev. Fluids}\ }\textbf {\bibinfo {volume} {5}},\ \bibinfo {pages}
  {044301} (\bibinfo {year} {2020})}\BibitemShut {NoStop}%
\bibitem [{\citenamefont {Prasad}\ \emph {et~al.}(2017)\citenamefont {Prasad},
  \citenamefont {Santangelo},\ and\ \citenamefont
  {Grason}}]{prasad2017subjamming}%
  \BibitemOpen
  \bibfield  {author} {\bibinfo {author} {\bibfnamefont {I.}~\bibnamefont
  {Prasad}}, \bibinfo {author} {\bibfnamefont {C.}~\bibnamefont {Santangelo}},\
  and\ \bibinfo {author} {\bibfnamefont {G.}~\bibnamefont {Grason}},\ }\bibinfo
  {title} {Subjamming transition in binary sphere mixtures},\ \href@noop {}
  {\bibfield  {journal} {\bibinfo  {journal} {Phys. Rev. E}\ }\textbf {\bibinfo
  {volume} {96}},\ \bibinfo {pages} {052905} (\bibinfo {year}
  {2017})}\BibitemShut {NoStop}%
\bibitem [{\citenamefont {Golick}\ and\ \citenamefont
  {Daniels}(2009)}]{golick2009mixing}%
  \BibitemOpen
  \bibfield  {author} {\bibinfo {author} {\bibfnamefont {L.~A.}\ \bibnamefont
  {Golick}}\ and\ \bibinfo {author} {\bibfnamefont {K.~E.}\ \bibnamefont
  {Daniels}},\ }\bibinfo {title} {Mixing and segregation rates in sheared
  granular materials},\ \href {https://doi.org/10.1103/PhysRevE.80.042301}
  {\bibfield  {journal} {\bibinfo  {journal} {Phys. Rev. E}\ }\textbf {\bibinfo
  {volume} {80}},\ \bibinfo {pages} {042301} (\bibinfo {year}
  {2009})}\BibitemShut {NoStop}%
\bibitem [{\citenamefont {Schlick}\ \emph {et~al.}(2015)\citenamefont
  {Schlick}, \citenamefont {Fan}, \citenamefont {Isner}, \citenamefont
  {Umbanhowar}, \citenamefont {Ottino},\ and\ \citenamefont
  {Lueptow}}]{schlick2015modeling}%
  \BibitemOpen
  \bibfield  {author} {\bibinfo {author} {\bibfnamefont {C.~P.}\ \bibnamefont
  {Schlick}}, \bibinfo {author} {\bibfnamefont {Y.}~\bibnamefont {Fan}},
  \bibinfo {author} {\bibfnamefont {A.~B.}\ \bibnamefont {Isner}}, \bibinfo
  {author} {\bibfnamefont {P.~B.}\ \bibnamefont {Umbanhowar}}, \bibinfo
  {author} {\bibfnamefont {J.~M.}\ \bibnamefont {Ottino}},\ and\ \bibinfo
  {author} {\bibfnamefont {R.~M.}\ \bibnamefont {Lueptow}},\ }\bibinfo {title}
  {Modeling segregation of bidisperse granular materials using physical control
  parameters in the quasi-2d bounded heap},\ \href@noop {} {\bibfield
  {journal} {\bibinfo  {journal} {AIChE J.}\ }\textbf {\bibinfo {volume}
  {61}},\ \bibinfo {pages} {1524} (\bibinfo {year} {2015})}\BibitemShut
  {NoStop}%
\bibitem [{\citenamefont {Marks}\ \emph {et~al.}(2012)\citenamefont {Marks},
  \citenamefont {Rognon},\ and\ \citenamefont {Einav}}]{marks2012grainsize}%
  \BibitemOpen
  \bibfield  {author} {\bibinfo {author} {\bibfnamefont {B.}~\bibnamefont
  {Marks}}, \bibinfo {author} {\bibfnamefont {P.}~\bibnamefont {Rognon}},\ and\
  \bibinfo {author} {\bibfnamefont {I.}~\bibnamefont {Einav}},\ }\bibinfo
  {title} {Grainsize dynamics of polydisperse granular segregation down
  inclined planes},\ \href@noop {} {\bibfield  {journal} {\bibinfo  {journal}
  {J. Fluid Mech.}\ }\textbf {\bibinfo {volume} {690}},\ \bibinfo {pages} {499}
  (\bibinfo {year} {2012})}\BibitemShut {NoStop}%
\bibitem [{\citenamefont {Tunuguntla}\ \emph {et~al.}(2014)\citenamefont
  {Tunuguntla}, \citenamefont {Bokhove},\ and\ \citenamefont
  {Thornton}}]{tunuguntla2014mixture}%
  \BibitemOpen
  \bibfield  {author} {\bibinfo {author} {\bibfnamefont {D.~R.}\ \bibnamefont
  {Tunuguntla}}, \bibinfo {author} {\bibfnamefont {O.}~\bibnamefont
  {Bokhove}},\ and\ \bibinfo {author} {\bibfnamefont {A.~R.}\ \bibnamefont
  {Thornton}},\ }\bibinfo {title} {A mixture theory for size and density
  segregation in shallow granular free-surface flows},\ \href@noop {}
  {\bibfield  {journal} {\bibinfo  {journal} {J. Fluid Mech.}\ }\textbf
  {\bibinfo {volume} {749}},\ \bibinfo {pages} {99} (\bibinfo {year}
  {2014})}\BibitemShut {NoStop}%
\bibitem [{\citenamefont {Powell}(1979)}]{powell1979site}%
  \BibitemOpen
  \bibfield  {author} {\bibinfo {author} {\bibfnamefont {M.~J.}\ \bibnamefont
  {Powell}},\ }\bibinfo {title} {Site percolation in randomly packed spheres},\
  \href@noop {} {\bibfield  {journal} {\bibinfo  {journal} {Phys. Rev. B}\
  }\textbf {\bibinfo {volume} {20}},\ \bibinfo {pages} {4194} (\bibinfo {year}
  {1979})}\BibitemShut {NoStop}%
\bibitem [{\citenamefont {Ziff}\ and\ \citenamefont
  {Torquato}(2017)}]{ziff2017percolation}%
  \BibitemOpen
  \bibfield  {author} {\bibinfo {author} {\bibfnamefont {R.~M.}\ \bibnamefont
  {Ziff}}\ and\ \bibinfo {author} {\bibfnamefont {S.}~\bibnamefont
  {Torquato}},\ }\bibinfo {title} {Percolation of disordered jammed sphere
  packings},\ \href@noop {} {\bibfield  {journal} {\bibinfo  {journal} {J.
  Phys. A Math. Theor.}\ }\textbf {\bibinfo {volume} {50}},\ \bibinfo {pages}
  {085001} (\bibinfo {year} {2017})}\BibitemShut {NoStop}%
\bibitem [{\citenamefont {He}\ and\ \citenamefont
  {Ekere}(2004)}]{he2004effect}%
  \BibitemOpen
  \bibfield  {author} {\bibinfo {author} {\bibfnamefont {D.}~\bibnamefont
  {He}}\ and\ \bibinfo {author} {\bibfnamefont {N.~N.}\ \bibnamefont {Ekere}},\
  }\bibinfo {title} {Effect of particle size ratio on the conducting
  percolation threshold of granular conductive--insulating composites},\
  \href@noop {} {\bibfield  {journal} {\bibinfo  {journal} {J. Phys. D}\
  }\textbf {\bibinfo {volume} {37}},\ \bibinfo {pages} {1848} (\bibinfo {year}
  {2004})}\BibitemShut {NoStop}%
\bibitem [{\citenamefont {Gray}\ and\ \citenamefont
  {Thornton}(2005)}]{gray2005theory}%
  \BibitemOpen
  \bibfield  {author} {\bibinfo {author} {\bibfnamefont {J.~M. N.~T.}\
  \bibnamefont {Gray}}\ and\ \bibinfo {author} {\bibfnamefont {A.~R.}\
  \bibnamefont {Thornton}},\ }\bibinfo {title} {A theory for particle size
  segregation in shallow granular free-surface flows},\ \href@noop {}
  {\bibfield  {journal} {\bibinfo  {journal} {Proc. R. Soc. A}\ }\textbf
  {\bibinfo {volume} {461}},\ \bibinfo {pages} {1447} (\bibinfo {year}
  {2005})}\BibitemShut {NoStop}%
\bibitem [{\citenamefont {Gajjar}\ and\ \citenamefont
  {Gray}(2014)}]{gajjar2014asymmetric}%
  \BibitemOpen
  \bibfield  {author} {\bibinfo {author} {\bibfnamefont {P.}~\bibnamefont
  {Gajjar}}\ and\ \bibinfo {author} {\bibfnamefont {J.~M. N.~T.}\ \bibnamefont
  {Gray}},\ }\bibinfo {title} {Asymmetric flux models for particle-size
  segregation in granular avalanches},\ \href@noop {} {\bibfield  {journal}
  {\bibinfo  {journal} {J. Fluid Mech.}\ }\textbf {\bibinfo {volume} {757}},\
  \bibinfo {pages} {297} (\bibinfo {year} {2014})}\BibitemShut {NoStop}%
\bibitem [{\citenamefont {Duan}\ \emph {et~al.}(2021)\citenamefont {Duan},
  \citenamefont {Umbanhowar}, \citenamefont {Ottino},\ and\ \citenamefont
  {Lueptow}}]{duan2021modelling}%
  \BibitemOpen
  \bibfield  {author} {\bibinfo {author} {\bibfnamefont {Y.}~\bibnamefont
  {Duan}}, \bibinfo {author} {\bibfnamefont {P.~B.}\ \bibnamefont
  {Umbanhowar}}, \bibinfo {author} {\bibfnamefont {J.~M.}\ \bibnamefont
  {Ottino}},\ and\ \bibinfo {author} {\bibfnamefont {R.~M.}\ \bibnamefont
  {Lueptow}},\ }\bibinfo {title} {Modelling segregation of bidisperse granular
  mixtures varying simultaneously in size and density for free surface flows},\
  \href@noop {} {\bibfield  {journal} {\bibinfo  {journal} {J. Fluid Mech.}\
  }\textbf {\bibinfo {volume} {918}},\ \bibinfo {pages} {A20} (\bibinfo {year}
  {2021})}\BibitemShut {NoStop}%
\bibitem [{\citenamefont {Tunuguntla}\ \emph {et~al.}(2017)\citenamefont
  {Tunuguntla}, \citenamefont {Weinhart},\ and\ \citenamefont
  {Thornton}}]{tunuguntla2017comparing}%
  \BibitemOpen
  \bibfield  {author} {\bibinfo {author} {\bibfnamefont {D.~R.}\ \bibnamefont
  {Tunuguntla}}, \bibinfo {author} {\bibfnamefont {T.}~\bibnamefont
  {Weinhart}},\ and\ \bibinfo {author} {\bibfnamefont {A.~R.}\ \bibnamefont
  {Thornton}},\ }\bibinfo {title} {Comparing and contrasting size-based
  particle segregation models},\ \href@noop {} {\bibfield  {journal} {\bibinfo
  {journal} {Comput. Part. Mech.}\ }\textbf {\bibinfo {volume} {4}},\ \bibinfo
  {pages} {387} (\bibinfo {year} {2017})}\BibitemShut {NoStop}%
\bibitem [{\citenamefont {Fan}\ and\ \citenamefont
  {Hill}(2011)}]{fan2011theory}%
  \BibitemOpen
  \bibfield  {author} {\bibinfo {author} {\bibfnamefont {Y.}~\bibnamefont
  {Fan}}\ and\ \bibinfo {author} {\bibfnamefont {K.~M.}\ \bibnamefont {Hill}},\
  }\bibinfo {title} {Theory for shear-induced segregation of dense granular
  mixtures},\ \href@noop {} {\bibfield  {journal} {\bibinfo  {journal} {New J.
  Phys.}\ }\textbf {\bibinfo {volume} {13}},\ \bibinfo {pages} {095009}
  (\bibinfo {year} {2011})}\BibitemShut {NoStop}%
\end{thebibliography}%
\bibliographystyle{apsrev4-2}
\end{document}